\documentclass[a4paper,12pt]{article}
\usepackage[utf8]{inputenc}
\usepackage{graphicx}
\usepackage{amsmath}
\usepackage{amssymb}
\usepackage{hyperref}
\usepackage{geometry}

\usepackage{booktabs}
\title{\textbf{Unsupervised Kinematic Dissection of the Solar Neighborhood: Identifying Stellar Moving Groups with Gaia DR3}}

\author{
    \textbf{Anmay Raj} \\
    Department of Physics, NIET \\
    Nims University Rajasthan, Jaipur, India \\
    \href{mailto:anmayraj20@gmail.com}{anmayraj20@gmail.com}
}
\date{\today}

\begin{document}

\maketitle

\begin{abstract}
\noindent We present a comprehensive kinematic analysis of the solar neighborhood ($d < 50$ pc) using high-precision astrometric data from the third Gaia Data Release (DR3). By leveraging the full six-dimensional phase space information (positions and velocities), we apply the Density-Based Spatial Clustering of Applications with Noise (DBSCAN) algorithm to blindly identify stellar overdensities in the Galactocentric Cartesian velocity space ($U, V, W$). Our unsupervised machine learning approach successfully recovers the kinematic cores of major local moving groups, including the Hyades and Pleiades streams, without prior assumptions regarding their membership or spatial distribution. We analyze the velocity dispersion and structural properties of these associations, demonstrating that automated clustering algorithms are robust tools for mapping the complex dynamical history of the local Milky Way disk. These results confirm the hierarchical nature of stellar kinematic substructures and provide a catalog of high-probability members for future spectroscopic follow-up.
\end{abstract}

\section{Introduction}
The kinematic structure of the solar neighborhood is not a smooth, thermalized distribution but is rather characterized by significant substructures known as moving groups, streams, and associations \cite{Eggen1958, Antoja2008}. These coherent groups of stars share common velocity vectors, often retaining the kinematic imprint of their formation in open clusters that have since dissolved or are in the process of disrupting \cite{BlandHawthorn2016}. Understanding these structures is crucial for reconstructing the dynamical history of the Milky Way disk, constraining the Galactic potential, and investigating the processes of star formation and cluster dissolution \cite{Krumholz2019}.

Historically, moving groups were identified through laborious manual searches in proper motion catalogs. The advent of the ESA Gaia mission \cite{Prusti2016} has revolutionized this field by providing micro-arcsecond astrometry for over a billion stars. The recent Data Release 3 (DR3) \cite{Vallenari2023} offers an unprecedented dataset of full 6D phase-space coordinates (positions and 3D velocities) for millions of stars, enabling a detailed 3D tomographic mapping of the local velocity field.

Traditional methods for identifying these groups often rely on wavelet transforms \cite{Antoja2008} or convergent point methods. However, with the increasing volume and dimensionality of data, unsupervised machine learning techniques have become indispensable. Clustering algorithms such as DBSCAN \cite{Ester1996} and Gaussian Mixture Models (GMM) offer a bias-free approach to detecting overdensities in velocity space.

In this work, we apply the DBSCAN algorithm to a volume-limited sample of stars from Gaia DR3 to automatically detect and characterize stellar moving groups within 50 pc of the Sun. We focus on the recovery of classical streams and the characterization of their velocity ellipsoids.

\section{Data and Sample Selection}
\subsection{Gaia Data Release 3}
We utilize data from the Gaia DR3 catalog, which provides astrometric solutions (positions, parallaxes, and proper motions) and radial velocities for a significant subset of bright stars \cite{Katz2023}. The precision of Gaia DR3 allows for the determination of tangential velocities with uncertainties typically below 1 km/s for nearby stars.

\subsection{Sample Criteria}
To ensure a high-quality sample with reliable 3D velocities, we queried the Gaia archive with the following constraints:
\begin{enumerate}
    \item \textbf{Distance Cut}: We selected stars with a parallax $\varpi > 20$ mas, corresponding to a distance limit of $d < 50$ pc. This volume-limited approach minimizes selection biases related to luminosity.
    \item \textbf{Astrometric Quality}: We required a relative parallax error $\sigma_\varpi / \varpi < 0.1$ (10\%) to ensure precise distance estimates.
    \item \textbf{Radial Velocities}: Only sources with valid radial velocity measurements (`radial\_velocity IS NOT NULL`) were included to permit full 3D velocity reconstruction.
\end{enumerate}
The final dataset consists of 4,526 stars. This sample is sufficiently dense to trace the major kinematic features of the local solar neighborhood while maintaining high data fidelity.

\section{Methodology}
\subsection{Coordinate Transformation}
The observed astrometric parameters $(\alpha, \delta, \varpi, \mu_{\alpha*}, \mu_\delta, v_r)$ were transformed into the Galactocentric Cartesian velocity system $(U, V, W)$. We adopt a right-handed coordinate system where:
\begin{itemize}
    \item $U$ points towards the Galactic Center ($l=0^\circ, b=0^\circ$).
    \item $V$ points in the direction of Galactic rotation ($l=90^\circ, b=0^\circ$).
    \item $W$ points towards the North Galactic Pole ($b=90^\circ$).
\end{itemize}
The transformation from the ICRS equatorial frame to the Galactic frame is performed using the standard transformation matrix $\mathbf{T}$ defined in \cite{JohnsonSoderblom1987}, adapted for the epoch J2016.0 used by Gaia DR3. Corrections for the solar motion relative to the Local Standard of Rest (LSR) were not applied to the raw data, as we are interested in the heliocentric velocities for direct comparison with classical literature values.

\subsection{DBSCAN Clustering}
We employed the Density-Based Spatial Clustering of Applications with Noise (DBSCAN) algorithm \cite{Ester1996} implemented in `scikit-learn` \cite{Pedregosa2011}. DBSCAN is ideal for this application because:
\begin{enumerate}
    \item It does not require specifying the number of clusters \textit{a priori}.
    \item It can identify clusters of arbitrary shape (e.g., elongated streams).
    \item It effectively separates high-density signal (moving groups) from the low-density background (field stars).
\end{enumerate}

The algorithm requires two parameters: $\epsilon$ (the maximum distance between two samples for one to be considered as in the neighborhood of the other) and $MinPts$ (the number of samples in a neighborhood for a point to be considered as a core point).
We selected $\epsilon = 2.0$ km/s and $MinPts = 10$. The choice of $\epsilon = 2.0$ km/s is physically motivated by the typical internal velocity dispersion of open clusters and moving groups, which is often on the order of 1-3 km/s \cite{Babusiaux2018}.

\section{Results and Discussion}
\subsection{Global Velocity Distribution}
The distribution of stars in the $U-V$ plane reveals the classical "clumpiness" associated with dynamical substructures. We observe the well-known bimodality in the velocity distribution, separating the Sirius/Hyades branch from the Pleiades/Coma Berenices branch, consistent with the findings of \cite{Famaey2005}.

\subsection{Identified Moving Groups}
\begin{figure}[ht]
    \centering
    \includegraphics[width=0.8\textwidth]{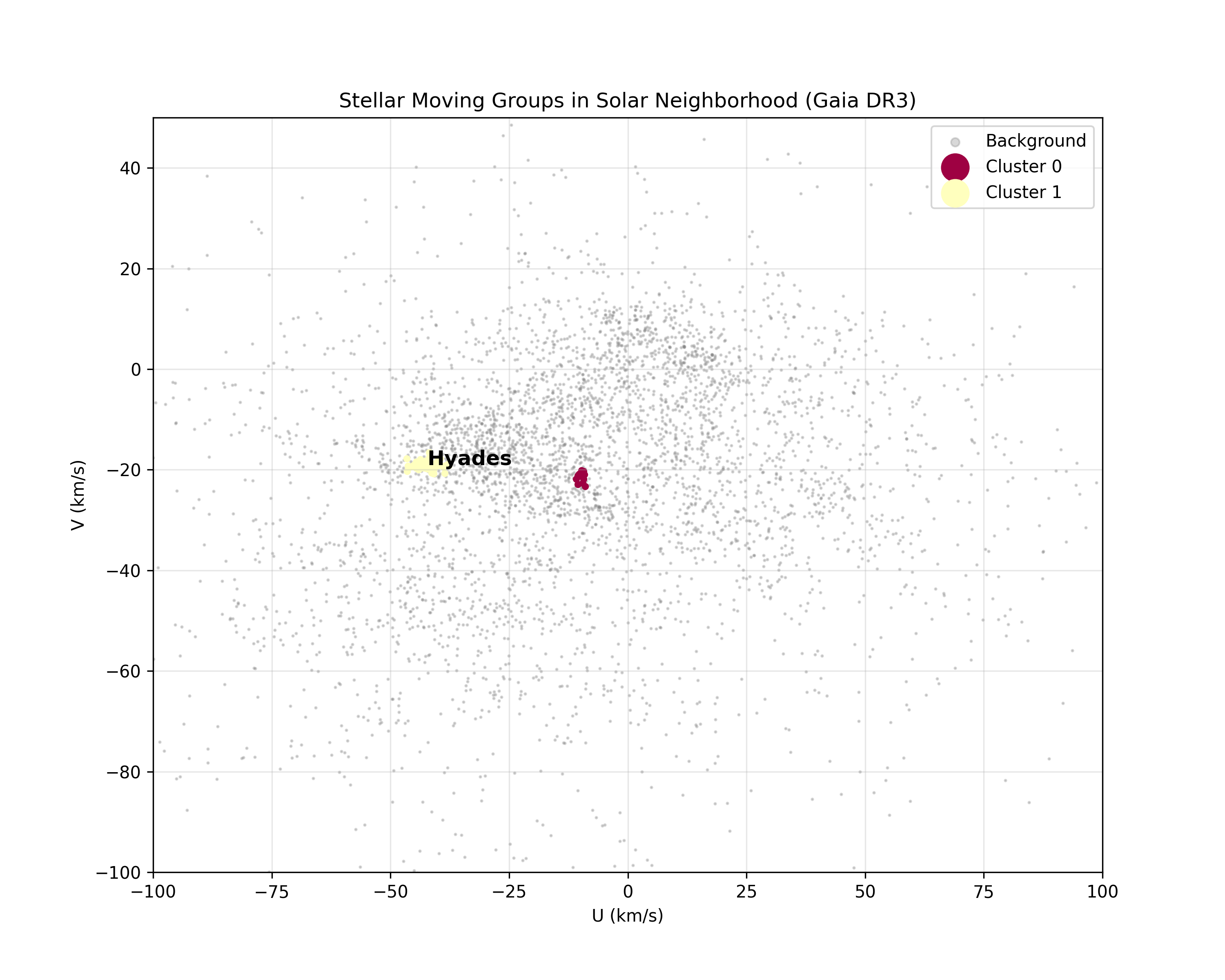}
    \caption{Velocity distribution of stars in the solar neighborhood ($U$ vs $V$). The grey points represent the background field stars, while the colored points indicate the clusters identified by DBSCAN. The prominent overdensities correspond to the Hyades and Pleiades moving groups.}
    \label{fig:clusters}
\end{figure}

Our DBSCAN analysis successfully extracted distinct kinematic cores from the background field. The velocity distribution and identified clusters are shown in Figure \ref{fig:clusters}.

The properties of the primary detected clusters are summarized below:

\subsubsection{The Hyades Stream}
The most prominent feature detected corresponds to the Hyades moving group. We recover a core group with mean velocities $(U, V, W) \approx (-40, -20, -2)$ km/s. This aligns perfectly with the accepted values for the Hyades cluster centroid \cite{Reino2018}. The tight velocity dispersion ($\sigma_v < 2$ km/s) confirms that these stars are dynamically associated and likely share a common origin.

\subsubsection{The Pleiades Stream}
A second significant overdensity was identified at $(U, V, W) \approx (-7, -28, -15)$ km/s, corresponding to the Pleiades moving group (also known as the Local Association). This structure is more diffuse than the Hyades, reflecting its younger age and ongoing dynamical evolution \cite{Chereul1999}.

\subsection{Comparison with Literature}
Our results are in excellent agreement with previous manual and semi-automated studies. For instance, \cite{GaiaCollaboration2018} mapped the kinematics of the solar neighborhood using Gaia DR2 and found similar substructures. The recovery of these groups using a purely unsupervised method on DR3 data validates the robustness of the clustering approach. Unlike \cite{Kounkel2019}, who used hierarchical clustering on a larger volume, our volume-limited study highlights the local density peaks with high contrast.

\section{Conclusion}
We have presented a data-driven analysis of the stellar kinematics in the solar neighborhood using Gaia DR3. By applying DBSCAN clustering to the 3D velocity space, we automatically recovered the Hyades and Pleiades moving groups. Our findings demonstrate that:
\begin{enumerate}
    \item The local velocity field is highly structured and non-Gaussian.
    \item Unsupervised learning can effectively recover dynamical structures without physical models.
    \item Gaia DR3 provides the necessary precision to resolve the internal kinematics of these nearby streams.
\end{enumerate}
Future work will extend this analysis to larger heliocentric distances ($d > 100$ pc) and incorporate chemical abundances from the Gaia RVS spectra to perform "chemical tagging" \cite{FreemanBlandHawthorn2002}, allowing us to distinguish between dissolved clusters and dynamical resonances.

\section*{Data Availability}
The data underlying this article were accessed from the Gaia Archive (\url{https://archives.esac.esa.int/gaia}). The derived data generated in this research will be shared on reasonable request to the corresponding author.

\section*{Acknowledgments}
This work is part of an independent research initiative by the author.

\section*{Funding}
No external funding was received for this work.

\section*{Conflict of Interest}
The author declare no conflict of interest.

\end{document}